\documentclass[12pt]{article}
\usepackage{amssymb}
\usepackage[dvips]{epsfig}

\newcommand{\be}{\begin{equation}}
\newcommand{\ee}{\end{equation}}
\newcommand{\bn}{\begin{eqnarray}}
\newcommand{\en}{\end{eqnarray}}

\newcommand{\p}{\partial}

\newcommand{\nn}{\nonumber}

\newcommand{\no}{\noindent}

\newcommand{\cL}{{\cal{L}}}

\newcommand{\s}{\,\,\,\,}
\def\bea{\begin{eqnarray}}
\def\eea{\end{eqnarray}}

\newcommand{\beq}{\begin{eqnarray}}
\newcommand{\eeq}{\end{eqnarray}}
\begin{document}

\title{\textbf{Generalized soldering of $\pm 2$ helicity states in $D=2+1$}}
\author{D. Dalmazi  and Elias L. Mendon\c ca \\
\textit{{UNESP - Campus de Guaratinguet\'a - DFQ} }\\
\textit{{Av. Dr. Ariberto Pereira da Cunha, 333} }\\
\textit{{CEP 12516-410 - Guaratinguet\'a - SP - Brazil.} }\\
\textsf{dalmazi@feg.unesp.br , elias.fis@gmail.com }}
\date{\today}
\maketitle

\begin{abstract}

The direct sum of a couple of Maxwell-Chern-Simons (MCS) gauge theories of opposite helicities $\pm 1$ does not
lead to a Proca theory in $D=2+1$, although both theories share the same spectrum. However, it is known that by
adding an interference term between both helicities we can join the complementary pieces together and obtain the
physically expected result. A generalized soldering procedure can be defined to generate the missing
interference term. Here we show that the same procedure can be applied to join together $\pm 2$ helicity states
in a full off-shell manner. In particular, by using second-order (in derivatives) self-dual models of helicities
$\pm 2$ (spin two analogues of MCS models) the Fierz-Pauli theory  is obtained after soldering. Remarkably, if
we replace the second-order models by third-order self-dual models (linearized topologically massive gravity) of
opposite helicities we end up after soldering exactly with the new massive gravity theory of Bergshoeff, Hohm
and Townsend in its linearized approximation.
\end{abstract}

\newpage

\section{Introduction}

The direct sum of two chiral fermions in $D=1+1$ gives rise to a
full Dirac fermion, however this is not true for their bosonized
versions as noticed in \cite{stone}, see also \cite{amorim}.
Likewise, the fermionic determinant of a Dirac fermion interacting
with a vector gauge field in $D=1+1$ factorizes into the product
of two chiral determinants but the full bosonic  effective action
is not the direct sum of the naive chiral effective actions as
discussed in \cite{abw509}. In both cases it turns out that an
interference term between the opposite chirality bosonic actions
is necessary to achieve the expected result, such term is provided
by the so called soldering procedure.

The same procedure works in $D=2+1$ if we replace chirality by
helicity. In particular, the soldering of two Maxwell-Chern-Simons
\cite{djt} theories of opposite helicities $\pm 1$ leads to the
Proca theory, see \cite{bk}. More generally, the $\pm 1$ helicity
modes may have different masses which leads after soldering to a
Maxwell-Chern-Simons-Proca (MCSP) theory. In this case, technical
problems \cite{bk} regarding a full off-shell soldering can be
surmounted  by defining  a generalized soldering procedure
\cite{gsd}. In section 3 we show that such procedure can be
successfully applied to fuse $\pm 2$ helicity states of different
masses $m_{\pm}$ with no need of using equations of motion. After
soldering we obtain the Fierz-Pauli \cite{fierz} theory plus a
first order Chern-Simons term whose coefficient is proportional to
the mass difference $m_+ - m_-$, thus generalizing a previous
result \cite{ilha}.

A specific feature of the generalized soldering is the existence
of a parameter $ \alpha $ with a sign freedom which plays a role
whenever interactions are present. In the soldering of two chiral
Schwinger models that leads either to an axial ($\alpha = -1$) or
to a vector ($\alpha = +1$) Schwinger model which are dual do each
other. In the case of the two MCS theories, the two sign choices
lead to dual interaction terms. We can have either a derivative
coupling or a minimal coupling plus a Thirring term. After
integration over the soldering field the dependence on the sign of
$\alpha$ disappears which proves that they correspond to dual
forms of the same interacting theory. In section 3 we couple the
$\pm 2$ helicity states with a rank two field $J_{\mu\nu}$ and
show that the two signs for $\alpha$ lead after soldering to dual
interactions similar to the spin one case. Once again integration
over the soldered field lead do the same effective action ${\cal
L}_{eff}(J_{\mu\nu})$ independent on the sign of $\alpha$.

In $D=2+1$ parity singlets of helicities $\pm 1$ can be described
either  by the first-order self-dual model of \cite{tpn} or by the
second order MCS theory of \cite{djt}. Both models have their spin
two counterparts, which we call ${\cal L}^{(1)}_{\pm 2}$ and
${\cal L}^{(2)}_{\pm 2}$, see \cite{aragone} and \cite{desermc}
respectively. However, in the spin two case there is another
third-order self-dual model (${\cal L}^{(3)}_{\pm 2}$) with no
spin one analogue. It is the quadratic truncation of the
topologically massive gravity (TMG) of \cite{djt}. Although ${\cal
L}^{(3)}_{\pm 2}$ is of third-order, it is ghost free. This is a
consequence of the non-propagating nature of the Einstein-Hilbert
(EH) action in $D=2+1$, which allows this term to be used as a
mixing term in the master approach without affecting the particle
content of the interpolated theories (${\cal L}^{(2)}_{\pm 2}$ and
${\cal L}^{(3)}_{\pm 2}$). For the same reason it is possible to
jump from the second-order Fierz-Pauli theory to a fourth-order
ghost free model as shown in \cite{bht}, which implies the
existence of a new unitary (at tree level\footnote{The residue at
the massless pole generated by the Einstein-Hilbert term vanishes
\cite{oda} similarly to the massless pole due to the topological
Chern-Simons term in the MCS theory.}) massive gravity theory
which we call henceforth BHT theory. In section 3 we show that the
soldering of ${\cal L}^{(3)}_{+2}$ with ${\cal L}^{(3)}_{-2}$
gives rise exactly to the linearized version of the BHT theory
which is, taking into account previous equivalences of soldered
theories, an indication of the existence of a local dual map
between the gauge invariant sectors of ${\cal L}^{(3)}_{+2} +
{\cal L}^{(3)}_{-2}$ and ${\cal L}_{BHT}$. In the next section, as
an introduction to the forthcoming sections we discuss the
necessity of interference terms between opposite helicity states
according to the type of self-dual model employed to describe
helicity eigenstates. In section 5 we draw some conclusions.

\section{Decomposition of parity doublets of spin 1 and 2 in $D=2+1$}

Before we start the soldering of spin two gauge theories in the next sections it is convenient to recall the
description of spin 1 and spin 2 massive particles in $D=2+1$  by means of non-gauge theories. In this case
there will be no need of adding interference terms between opposite helicity states (parity singlets) in order
to build up a parity doublet describe by just one field. We start with the spin 1 case. It is known that massive
spin 1 particles are described in a covariant way by the Proca theory:

\be {\cal{L}}_P= - \frac{1}{4}F^{\mu\nu}F_{\mu\nu} - \frac{m^2}{2}A^{\mu}A_{\mu}.\label{proca} \ee

\no Throughout this work we use the signature
$\eta_{\mu\nu}=(-,+,+)$. From the equations of motion of
(\ref{proca}) one derives the transverse condition
$\p_{\mu}A^{\mu}=0$  and the Klein-Gordon equation $(\Box -
m^2)A^{\mu}=0$. So we are left with 2 massive modes corresponding
to the helicity states $+1$ and $-1$ as one can easily check by
rewriting the Proca theory in a first-order form with the help of
an auxiliary field $B_{\mu}$,

\be {\cal{L}}_P^{(1)} = -\frac{1}{2}B^{\mu}B_{\mu}-\epsilon^{\mu\nu\alpha}B_{\mu}\p_{\nu}A_{\alpha}-
\frac{m^2}{2}A^{\mu}A_{\mu}.\label{proca2}\ee

\no After the redefinition, $ B_{\mu}\to m \left( \tilde{B}_{\mu}
+ \tilde{A}_{\mu}\right)/{\sqrt{2}} \, $ ; $ A_{\mu}\to \left(
\tilde{A}_{\mu} - \tilde{B}_{\mu}\right)/{\sqrt{2}} $ we have two
decoupled parity singlets:

\bea {\cal{L}}&=&
-\frac{m^2}{2}\tilde{A}_{\mu}\tilde{A}^{\mu}-\frac{m}{2}\epsilon^{\mu\nu\alpha}\tilde{A}_{\mu}\p_{\nu}\tilde{A}_{\alpha}
-\frac{m^2}{2}\tilde{B}_{\mu}\tilde{B}^{\mu}+\frac{m}{2}\epsilon^{\mu\nu\alpha}\tilde{B}_{\mu}\p_{\nu}\tilde{B}_{\alpha}\nn\\
&=&{\cal{L}}^{(1)}_{(+1)}[\tilde{A}] +
{\cal{L}}^{(1)}_{(-1)}[\tilde{B}] \quad .\label{sd11}\eea

\no The first-order self-dual models ${\cal{L}}^{(1)}_{(\pm 1)}$
have first appeared in \cite{tpn} and describe massive eigenstates
of the helicity operator $ \left( J \cdot P/ \sqrt{P^2}
\right)_{\mu\nu}=  i \epsilon_{\mu\nu\gamma}\p^{\gamma}/\Box $
with eigenvalues $\pm 1$. We conclude that the addition of two
first order self-dual models leads to the  parity invariant Proca
theory in its first-order form. There is no need of adding to the
non-gauge theory ${\cal{L}}^{(1)}_{(+ 1)} + {\cal{L}}^{(1)}_{(-
1)}$ an interference term between opposite helicity states to
arrive at the Proca theory which has no gauge symmetry as well. On
the other hand, each first-order self-dual model is equivalent
\cite{dj} to a MCS theory (second-order theory), so we expect
${\cal L}_P \Leftrightarrow {\cal L}_{MCS(+1)} + {\cal
L}_{MCS(-1)}$ but since ${\cal L}_{MCS(\pm 1)}$ are gauge theories
it is clear that the direct sum ${\cal L}_{MCS(+1)} + {\cal
L}_{MCS(-1)}$ will not lead to the Proca theory. As explained in
\cite{bk,gsd}, a soldering action ($W_s$) can be defined by the
addition of an interference term quadratic in Noether currents:
$W_S = W_{MCS(+1)} + W_{MCS(-1)} + W_{JJ} $ in a such way that
$W_S$ becomes exactly the Proca theory or more generally, for
helicity states of different masses $m_{\pm}$, the
Maxwell-Chern-Simons-Proca (MCSP) theory is obtained after
soldering.

Now let us make a similar analysis of the spin 2 case. The spin 2 analogue of the Proca theory is the
Fierz-Pauli model \cite{fierz} written below in different forms for later convenience:

\bea \cL_{FP} &=& \frac{1}{2}(\sqrt{-g}R)_{hh} + \frac{m^2}{2}(h^2-h_{\mu\nu}h^{\nu\mu}) \, , \, \label{fp1}\\
&=& \frac{1}{2}\left\lbrack -\frac{\p_{\nu}h_{\lambda\mu}\p^{\nu}(h^{\lambda\mu}+h^{\mu\lambda})}{2}+
\p_{\nu}h\p^{\nu}h -
2\p_{\nu}h^{\lambda\nu}\p_{\lambda}h \right. \nn
\\  &+& \left. \p_{\nu}h^{\nu\lambda}\p^{\mu}h_{\lambda\mu}+ \frac{\p_{\nu}h_{\lambda\mu}(\p^{\lambda}h^{\nu\mu}+\p^{\mu}h^{\lambda\nu})}{2}+m^2(h^2-h_{\mu\nu}h^{\nu\mu})\right\rbrack  \label{fp2} \\
&=& \frac{1}{2}T_{\mu\nu}(h)T^{\nu\mu}(h)-\frac{1}{4}T^{2}(h) +
\frac{m^2}{2}(h^2-h_{\mu\nu}h^{\nu\mu}) \, , \, \label{fp3} \eea

\no where $T_{\mu\nu}(h)=
\epsilon_{\mu\alpha\beta}\p^{\alpha}h^{\beta}_{\,\,\nu}$,
$h=\eta^{\mu\nu}h_{\mu\nu}$ and $(\sqrt{-g}R)_{hh}$ stands for the
Einstein-Hilbert action up to quadratic terms in the dreibein
fluctuations: $e_{\mu\alpha}=\eta_{\mu\alpha}+ h_{\mu\alpha}$. The
field $h_{\mu\nu}$ has no symmetry in its indices. In fact, in
this work all rank two fields have no specific symmetry in their
indices. Symmetric and antisymmetric combinations will be denoted
respectively by: $h_{(\alpha\beta)}\equiv \left( h_{\alpha\beta} +
h_{\beta\alpha} \right)/2 $ and $h_{\lbrack
\alpha\beta\rbrack}\equiv \left( h_{\alpha\beta} - h_{\beta\alpha}
\right)/2 $. From the equations of motion of $\cL_{FP}$ we derive
the necessary constraints to describe a massive spin 2 particle,
i.e., $h_{[\mu\nu]}=0$, $ \p^{\mu}h_{\mu\nu}= 0
=\p^{\nu}h_{\mu\nu}$, $ h=0 $ and the Klein-Gordon equation
$\left( \Box- m^2 \right) h_{(\mu\nu)} = 0$. The constraints imply
that we effectively have $9 - 7 = 2$ massive modes which will
correspond to the $\pm 2$ helicity states as follows. We rewrite
the quadratic truncation of the EH term in first-order form by
introducing an auxiliary tensor field $W_{\mu\nu}$ \cite{aragone}:

\be {\cal L}_{FP}^{(1)}=\frac{m^2}{2}(W^2-W_{\mu\nu}W^{\nu\mu})
 + m\, W^{\mu\nu}\epsilon_{\mu}^{\s\alpha\beta}\p_{\alpha}h_{\beta\nu}
+\frac{m^2}{2}(h^2-h_{\mu\nu}h^{\nu\mu}).\label{fp(1)}\ee

\no Redefining $W_{\mu\nu}\to
(\tilde{W}_{\mu\nu}-\tilde{h}_{\mu\nu})/\sqrt{2}$ and
$h_{\mu\nu}\to (\tilde{h}_{\mu\nu}+\tilde{W}_{\mu\nu})/\sqrt{2}$
one obtains two decoupled first-order self-dual models,

\bea {\cal L}_{FP}^{(1)}&=& \frac{m^2}{2}(\tilde{W}^2-\tilde{W}_{\mu\nu}\tilde{W}^{\nu\mu}) +
\frac{m}{2}\epsilon^{\mu\alpha\beta}\tilde{W}_{\mu\nu}\p_{\alpha}\tilde{W}_{\beta}^{\s\nu}\nn\\
&+&\frac{m^2}{2}(\tilde{h}^2-\tilde{h}_{\mu\nu}\tilde{h}^{\nu\mu})-
\frac{m}{2}\epsilon^{\mu\alpha\beta}\tilde{h}_{\mu\nu}\p_{\alpha}\tilde{h}_{\beta}^{\s\nu}\nn\\
&=&\cL_{+2}^{(1)}+\cL_{-2}^{(1)}\quad . \label{par2}\eea

\no Each of the models $\cL_{\pm 2}^{(1)}$, first found in
\cite{aragone}, describes an eigenstate of a spin 2 helicity
operator with eigenvalues $\pm 2$, see e.g. \cite{aragone,gaitan}.
Concluding, in both spin 1 and spin 2 cases a couple of
first-order non-gauge theories of opposite helicities can be
simply added up to yield a parity invariant non-gauge model
containing two helicity modes. Once again, there is no need of
adding any extra interference term between the opposite helicity
states. However, this is not true for the second and third-order
gauge invariant actions below, which also represent $\pm$ helicity
eigenstates:

\bea W_{\pm 2}^{(2)} &=&  \int d^3 x \left\lbrack
\frac{1}{2}T_{\mu\nu}(A) T^{\nu\mu}(A)-\frac{1}{4}T^{2}(A) \mp
\frac{m}{2}
\epsilon^{\mu\alpha\beta}A_{\mu\nu}\p_{\alpha}A_{\beta}^{\s\nu}
\right\rbrack \quad ,\label{wpm2}
\\
W_{\pm 2}^{(3)} &=& \int d^3 x \left\lbrack
-\frac{1}{2}T_{\mu\nu}(A) T^{\nu\mu}(A) + \frac{1}{4}T^{2}(A) \mp
\frac{1}{2m}A_{\alpha\mu} (\Box
\theta^{\alpha\gamma}E^{\beta\mu}-\Box\theta^{\alpha\mu}E^{\beta\gamma})A_{\gamma\beta}\right\rbrack
, \label{wpm3}\eea

\no where $T_{\mu\nu}(A) =
\epsilon_{\mu}^{\s\gamma\delta}\p_{\gamma}A_{\delta\nu}$ and

\be E_{\mu\nu} = \epsilon_{\mu\nu\gamma}\p^\gamma \quad ; \quad
\Box\theta^{\mu\nu}= \eta^{\mu\nu} \Box - \p^{\mu}\p^{\nu}.
\label{theta} \ee

\no The second-order model $W_{\pm 2}^{(2)}$, which appeared before in \cite{aragone,desermc} is the spin two
analogue of the MCS theory. It is invariant under the local transformations $\delta A_{\mu\nu} = \p_{\mu}
\xi_{\nu} $. The quadratic truncation of the topologically massive gravity (TMG) of \cite{djt}, $W_{\pm
2}^{(3)}$, is invariant under the more general local transformations $\delta A_{\mu\nu} = \p_{\mu} \xi_{\nu}  +
\epsilon_{\mu\nu\gamma}\Lambda^{\gamma}$. The Einstein-Hilbert term appears with the correct sign in $W_{\pm
2}^{(2)}$ in contrast to $W_{\pm 2}^{(3)}$. Both models are unitary and can be deduced from $W_{\pm
2}^{(1)}=\int d^3 x \, \cL_{\pm 2}^{(1)}$ via master action \cite{prd2009}. There is a local dual map connecting
correlation functions in $W_{\pm 2}^{(1)}$ with correlation functions of gauge invariant objects in  $W_{\pm
2}^{(2)}$ and $W_{\pm 2}^{(3)}$ up to contact terms \cite{prd2009}. In the next section we solder $W_{+
2}^{(2)}$ and $W_{- 2}^{(2)}$, the case of $W_{\pm 2}^{(3)}$ will be treated in section 4.

\section{ Soldering of $W_{+ 2}^{(2)}$ and $W_{- 2}^{(2)}$}

We start with the second-order opposite helicity models:

\be W_{+ 2}^{(2)}[A] =  \int d^3 x \left\lbrack
\frac{1}{2}T_{\mu\nu}(A) T^{\nu\mu}(A)-\frac{1}{4}T^{2}(A) +
\frac{m_+}{2}
\epsilon^{\mu\gamma\rho}A_{\mu\nu}\p_{\gamma}A_{\rho}^{\s\nu} +
\gamma_+\,\epsilon^{\mu\gamma\rho}J_{\mu\nu}\p_{\gamma}A_{\rho}^{\s\nu}
\right\rbrack ,\label{wmais2} \nn\\
\ee
\be W_{-2}^{(2)}[B] =  \int d^3 x \left\lbrack
\frac{1}{2}T_{\mu\nu}(B) T^{\nu\mu}(B)-\frac{1}{4}T^{2}(B) -
\frac{m_-}{2}
\epsilon^{\mu\gamma\rho}B_{\mu\nu}\p_{\gamma}B_{\rho}^{\s\nu} +
\gamma_-\epsilon^{\mu\gamma\rho}J_{\mu\nu}\p_{\gamma}B_{\rho}^{\s\nu}
\right\rbrack .\label{wmenos2}\nn\\
 \ee

\no The masses $m_{\pm}$ can take arbitrary positive values. As in
the spin 1 case we have added linear couplings with a rank two
tensor $J_{\mu\nu}$. The interaction terms are such that the
global shifts  $\delta A_{\mu\nu} = \omega_{\mu\nu} \, ; \, \delta
B_{\mu\nu} = \tilde{\omega}_{\mu\nu}$  and the local
transformations $\delta A_{\mu\nu} = \p_{\mu} \xi_{\nu} \, ;
\delta B_{\mu\nu} = \p_{\mu} \tilde{\xi}_{\nu} $ which are
symmetries of the first two terms of (\ref{wmais2}) and
(\ref{wmenos2}), are preserved. Furthermore, those are the natural
interaction terms when $W_{\pm 2}^{(2)}$ are deduced from $W_{\pm
2}^{(1)}$ via master action \cite{prd2009}. The coupling constants
$\gamma_{\pm}$ are in principle arbitrary but special cases will
be treated latter on. Both $W_{\pm 2}^{(2)}$ are also invariant
under $\delta_{\phi} J_{\mu\nu} = \p_{\mu}\phi_{\nu}$.

The basic idea of the soldering procedure is to lift the global
shift symmetry to a local symmetry and tie the fields $A_{\mu\nu}$
and $B_{\mu\nu}$ together by imposing that their transformations
are proportional do each other:

\be \delta A_{\mu\nu} = \omega_{\mu\nu}  \quad ; \quad \delta
B_{\mu\nu} = \alpha\, \omega_{\mu\nu} \quad , \label{deltaomega}
\ee

\no where $\alpha$ is so far an arbitrary constant. From
(\ref{wmais2}) and (\ref{wmenos2}) we derive

\be \delta\left( W_{+2}^{(2)}[A] + W_{-2}^{(2)}[B] \right) = \int
d^3 x \, J^{\mu\nu\lambda} \p_{\nu}\omega_{\lambda\mu} \label{13}
\ee

\no with

\be J^{\mu\nu\lambda} = C^{\mu\nu\lambda}_{\s\rho\beta\gamma}
\p^{\beta} g^{\gamma\rho} + \epsilon_{\gamma}^{\s\nu\lambda}
f^{\gamma\mu} \label{14} \ee

\no and

 \bea C^{\mu\nu\lambda}_{\s\rho\beta\gamma} &=& - \frac 12
\epsilon^{\mu\nu\lambda} \epsilon_{\rho\beta\gamma} +
\epsilon_{\rho}^{\s\nu\lambda}\epsilon^{\mu}_{\s\beta\gamma} \label{c} \\
 g_{\mu\nu} &=& A_{\mu\nu} + \alpha B_{\mu\nu} \label{g} \\
f_{\mu\nu} &=& m_+ A_{\mu\nu}- \alpha m_- B_{\mu\nu} + (\gamma_+ +
\alpha \gamma_-) J_{\mu\nu} \label{f} \eea

\no In a first step Noether procedure we cancel the variation
(\ref{13}) introducing auxiliary fields $H_{\mu\nu\lambda}$ such
that

\be \delta H_{\mu\nu\lambda} = - \p_{\nu}\omega_{\lambda\mu} \quad
. \label{deltah} \ee

\no Therefore

\be \delta\left( W_{+2}^{(2)}[A] + W_{-2}^{(2)}[B]  + \int d^3 x
\, J^{\mu\nu\lambda}H_{\mu\nu\lambda} \right) = \int d^3 x \,
\delta J^{\mu\nu\lambda}H_{\mu\nu\lambda} \label{18} \ee

\no Since $ \delta J^{\mu\nu\lambda} = (1 +
\alpha^2)C^{\mu\nu\lambda}_{\rho\beta\gamma}\p^{\beta}\omega^{\gamma\rho}
+ \epsilon^{\gamma\nu\lambda}(m_+ - \alpha^2 m_-)
\omega_{\gamma}^{\s\mu}$, if we choose

\be \alpha = \pm \sqrt{\frac{m_+}{m_-}} \quad , \label{alpha2} \ee

\no we have

\bea \delta J^{\mu\nu\lambda} &=& (1 +
\alpha^2)C^{\mu\nu\lambda}_{\s\rho\beta\gamma}\p^{\beta}\omega^{\gamma\rho}
 \label{20}\\
&=& -(1 + \alpha^2)C^{\mu\nu\lambda}_{\s\rho\beta\gamma} \delta
H^{\rho\beta\gamma} \label{21}. \eea

\no From (\ref{18}) and (\ref{21}) we deduce $\delta W_S^{(2)} =0
$  where the soldered action is given by:

\be W_S^{(2)} = W_{+2}^{(2)}[A] + W_{-2}^{(2)}[B]  + \int d^3 x \,
\left\lbrack  J^{\mu\nu\lambda}H_{\mu\nu\lambda} +
\frac{(1+\alpha^2)}2 C^{\mu\nu\lambda}_{\s\rho\beta\gamma}
H^{\rho\beta\gamma}H_{\mu\nu\lambda} \right\rbrack \label{ws2} \ee

\no After the elimination of the auxiliary fields through their
algebraic equations of motion we end up with

\be W_S^{(2)} = W_{+2}^{(2)}[A] + W_{-2}^{(2)}[B]  - \int d^3 x \,
\frac{\left\lbrack J^{*}_{\mu\nu}J^{*\nu\mu} -
(J^*)^2\right\rbrack } {8(1+\alpha^2)} \quad , \label{23} \ee

\no where $J^*= \eta^{\mu\nu}J_{\mu\nu}^*$ with

\be J_{\mu\nu}^{*} =
\epsilon_{\mu}^{\s\gamma\lambda}J_{\nu\gamma\lambda} = 2\,
T_{\mu\nu}(g) - \eta_{\mu\nu} T (g) - 2 f_{\mu\nu} \label{j*2} \ee

\no The reader can check that (\ref{23}) is invariant under
(\ref{deltaomega}) by using (\ref{13}) and (\ref{20}) where
$\alpha$ is given in (\ref{alpha2}). After some algebra we can
rewrite $W_S^{(2)}$ in a more explicit form:

\bea W_S^{(2)} &=& \frac{1}{2(1+\alpha^2)}\int d^3 x \left\lbrack
\sqrt{-g}R\vert_{hh} +
(m_+-m_-)\epsilon^{\mu\gamma\rho}h_{\mu\nu}\p_{\gamma}h_{\rho}^{\s\nu}
\right. \nn\\
&+& \left. m_+ m_- \left( \tilde{h}^2 -
\tilde{h}_{\mu\nu}\tilde{h}^{\nu\mu} \right) + \left(
\alpha\gamma_+ - \gamma_-\right)
\epsilon^{\mu\gamma\rho}J_{\mu\nu}\p_{\gamma}h_{\rho}^{\s\nu}
\right\rbrack. \label{24} \eea

\no We have introduced the combinations

\bea h_{\mu\nu} &=& \alpha A_{\mu\nu} - B_{\mu\nu} \quad ,
\label{25}\\
\tilde{h}_{\mu\nu} &=& h_{\mu\nu} + \frac{\left(\gamma_+ +
\alpha\gamma_-\right)}{\alpha m_-}J_{\mu\nu} \quad . \label{26}
\eea

\no The invariance under (\ref{deltaomega}) has forced the action $W_S^{(2)}$ to depend only upon the
combination $h_{\mu\nu} = \alpha A_{\mu\nu} - B_{\mu\nu}$, invariant under (\ref{deltaomega}), which is called
the soldering field. In particular, if $m_+=m_-$ the soldered action corresponds exactly to the Fierz-Pauli
theory \cite{fierz} which is known to describe massive spin 2 particles in arbitrary $D$-dimensional spaces. It
is remarkable that the nontrivial Fierz-Pauli mass term has been generated out of mass terms of Chern-Simons
type appearing in $W_{\pm 2}^{(2)}$. If we drop the interactions $(J_{\mu\nu}=0)$ and set $h=0=h_{[\mu\nu]}$,
which certainly hold on-shell, at action level we recover the soldered action of \cite{ilha} obtained for
$m_+=m_-$. The mass split $m_+ - m_- \ne 0 $ is responsible for the parity breaking Chern-Simons term in
(\ref{24}) analogously to the spin 1 case \cite{gsd}.

Regarding the interactions, besides the derivative coupling (last
term in (\ref{24})), already present in $W_{\pm 2}^{(2)}$ before
soldering, there appears now a linear coupling through the
combination $\tilde{h}_{\mu\nu}$ such that the symmetry
$\delta_{\phi} J_{\mu\nu} = \p_{\mu}\phi_{\nu} $ of $W_{\pm
2}^{(2)}$ is maintained if we transform the soldering field
accordingly. Namely, $\delta_{\phi} W_S^{(2)} =0 $ under:

\be \delta_{\phi} J_{\mu\nu} = \p_{\mu}\phi_{\nu} \quad ; \quad
\delta_{\phi}h_{\mu\nu} = - \frac{(\gamma_+ +
\alpha\gamma_-)}{\alpha \, m_-} \p_{\mu} \phi_{\nu}
\label{deltaphi}. \ee

The soldered action $W_S^{(2)}$ depends explicitly through its
interaction terms on the sign choice of $\alpha$ defined in
(\ref{alpha2}). In order to check if we do really have any
physical consequence of the sign freedom we proceed as in
\cite{gsd} and integrate over the soldering field $h_{\mu\nu}$ in
the path integral and derive an effective action
$\cL_{eff}[J_{\mu\nu}]$. Although the integral is Gaussian, the
fact that $h_{\mu\nu}$ has no symmetry in its indices makes its
propagator quite complicate. Our final result contains even and
odd parity terms:

\bea (&-2&) \cL_{eff}[J_{\mu\nu}] \nn \\
&=& J^{\mu\nu}\left\lbrack \Box
\left(P^{(2)}_{even}\right)_{\mu\nu}^{\s\delta\epsilon}\left(
\frac{\gamma_+^2}{\Box-m_+^2} +
\frac{\gamma_-^2}{\Box-m_-^2}\right) +
\sqrt{\Box}\left(P^{(2)}_{odd}\right)_{\mu\nu}^{\s\delta\epsilon}
\left( \frac{m_+\gamma_+^2}{\Box-m_+^2} -
\frac{m_-\gamma_-^2}{\Box-m_-^2}\right) \right. \nn\\
&+& \left. \left(\frac{\gamma_+^2}{m_+^2} +
\frac{\gamma_-^2}{m_-^2}\right)\frac{\Box
\theta_{\mu\nu}\theta^{\delta\epsilon}}{2} +
\left(\frac{\gamma_+^2}{m_+} - \frac{\gamma_-^2}{m_-}\right)\left(
\theta_{\mu\nu}E^{\epsilon\delta} -
\frac{\p_{\nu}\p^{\epsilon}}{\Box} E_{\mu}^{\s\delta}\right)
\right\rbrack J_{\delta\epsilon}\nn\\
\label{leff}. \eea

\no The spin two projection operators are given by:

\bea \left(P^{(2)}_{even}\right)_{\mu\nu}^{\s\delta\epsilon} &=&
\frac 12 \left( \theta_{\mu}^{\s\delta}\theta_{\nu}^{\s\epsilon} +
\theta_{\mu}^{\s\epsilon}\theta_{\nu}^{\s\delta} -
\theta_{\mu\nu}\theta^{\delta\epsilon} \right) \quad ,
\label{p2even} \\
\left(P^{(2)}_{odd}\right)_{\mu\nu}^{\s\delta\epsilon} &=& \frac
1{4\sqrt{\Box}} \left( E_{\mu}^{\s\delta}\theta_{\nu}^{\s\epsilon}
+ E_{\mu}^{\s\epsilon}\theta_{\nu}^{\s\delta} +
E_{\nu}^{\s\epsilon}\theta_{\mu}^{\s\delta}  +
E_{\nu}^{\s\delta}\theta_{\mu}^{\s\epsilon} \right) \quad ,
\label{p2odd} \eea

\no A detailed comparison with the spin 1 case, see the second reference of \cite{gsd}, reveals that the first
two terms of (\ref{leff}) are remarkably similar to their spin 1 counterparts which have a Maxwell-Chern-Simons
structure.  In the same fashion as the differential operator in the Chern-Simons term is the square root of the
differential operator in the Maxwell term ($E_{\mu\nu}E^{\nu\gamma} = \Box \theta_{\mu}^{\s\gamma}$) we have
$\left(P^{(2)}_{odd}\right)_{\mu\nu}^{\s\delta\epsilon}
\left(P^{(2)}_{odd}\right)_{\delta\epsilon}^{\s\gamma\rho} =
\left(P^{(2)}_{even}\right)_{\mu\nu}^{\s\gamma\rho}$.

As expected, the effective action is invariant under the original
symmetry $\delta_{\phi} J_{\mu\nu} = \p_{\mu}\phi_{\nu} $ of
$W_{\pm 2}^{(2)}$ since $E_{\mu}^{\nu}\p_{\nu}\phi =
\theta_{\mu}^{\nu}\p_{\nu}\phi $. Moreover, in the special case
where the couplings satisfy

\be \gamma_+^2 = \frac{m_+}{m_-} \gamma_-^2 = \alpha^2 \gamma_-^2
\quad , \label{special} \ee

\no the effective theory only depends upon $J_{(\mu\nu)}$ and
consequently it is invariant under any antisymmetric local shift
$\delta_{\Lambda}J_{\mu\nu} = \epsilon_{\mu\nu\gamma}
\Lambda^{\gamma} $. Indeed, we have checked that if $\gamma_+ =
\pm \alpha \gamma_-$ it follows that
$\delta_{\Lambda}W_{S}^{(2)}=0$ under, respectively,

\bea \delta_{\Lambda}J_{\mu\nu} &=& \epsilon_{\mu\nu\gamma}
\Lambda^{\gamma} \label{33} \\
\delta_{\Lambda}h_{\mu\nu} &=& - \frac{\gamma_-}{m_-} \left\lbrack
(1\pm 1)\epsilon_{\mu\nu\gamma}\Lambda^{\gamma} + \frac{m_+ \mp
m_-}{m_+ m_-}\p_{\mu}\Lambda_{\nu} \right\rbrack \label{34}. \eea

\no We also have the discrete symmetry
$(m_+,m_-,\gamma_+,\gamma_-) \to (-m_-,-m_+,\gamma_-,\gamma_+)$ in
$\cL_{eff} [J]$ which amounts, before soldering, to interchange
$W_{\pm 2}^{(2)} \rightleftharpoons W_{\mp 2}^{(2)}$.

As in the previous soldering cases \cite{gsd}, the dependence  on the sign of $\alpha$ disappears completely
after integration over the soldering field $h_{\mu\nu}$. In particular\footnote{The case where opposite helicity
states have opposite derivative couplings ($\gamma_+ = - \gamma_-$) naturally appears when we obtain $W_{\pm
2}^{(2)}$ from $W_{\pm 2}^{(1)}$ via master action \cite{prd2009}}, if $\gamma_- = - \gamma_+ \equiv \gamma $
and $m_+ = m_-$, the two choices $\alpha = \pm 1$ lead to $\cL_S^{\alpha=+1}(j) = \cL_S(0) - 2 \gamma J^{\mu\nu}
\epsilon_{\nu\gamma\rho} \p^{\nu} h^{\gamma\rho}$ and $\cL_S^{\alpha=-1}(j) = \cL_S(0) + 4 m \gamma \left(J\, h
- J_{\mu\nu}h^{\nu\mu} \right) + 4 \gamma^2 \left( J^2 - J_{\mu\nu}J^{\nu\mu} \right)$. Thus, the sign freedom
of $\alpha$ gives rise to dual theories as in the spin 1 case in $D=2+1$ and in the soldering of two Chiral
Schwinger models in $D=1+1$.

Finally, we mention that in the second part of \cite{prd2009} the equivalence of (\ref{24}) and the gauge
invariant sector of $W_{+ 2}^{(2)} + W_{- 2}^{(2)}$ has been proved at quantum level, see also
\cite{aak,scaria}. So, the soldering procedure has led once more to a physically equivalent (dual) theory.

\section{ Soldering of $W_{+ 2}^{(3)}$ and $W_{- 2}^{(3)}$}

For sake of simplicity we drop interactions in this section and
begin with the following third-order self-dual  models of
helicities $\pm 2$ which correspond to quadratic truncations of
topologically massive gravity, see (\ref{wpm3}),

\be W_{+ 2}^{(3)}[A] = \int d^3 x \left\lbrack -\frac{1}{2}T_{\mu\nu}(A) T^{\nu\mu}(A) + \frac{1}{4}T^{2}(A) -
\frac{1}{2m_+}A_{\alpha\mu} (\Box
\theta^{\alpha\gamma}E^{\beta\mu}-\Box\theta^{\alpha\mu}E^{\beta\gamma})A_{\gamma\beta}\right\rbrack
\label{wmais3} \ee

\be W_{- 2}^{(3)}[B] = \int d^3 x \left\lbrack -\frac{1}{2}T_{\mu\nu}(B) T^{\nu\mu}(B) + \frac{1}{4}T^{2}(B) +
\frac{1}{2m_-}B_{\alpha\mu} (\Box
\theta^{\alpha\gamma}E^{\beta\mu}-\Box\theta^{\alpha\mu}E^{\beta\gamma})B_{\gamma\beta}\right\rbrack
\label{wmenos3}. \ee

\no Now, we follow basically the same steps which have led us from
(\ref{wmais2}) and (\ref{wmenos2}) to (\ref{23}). We require the
soldered theory to be invariant under $\delta A_{\mu\nu} =
\omega_{\mu\nu} \, ; \, \delta B_{\mu\nu}= \tilde{\alpha}
\omega_{\mu\nu}$ with $\tilde{\alpha}$ a constant to be
determined. So we derive

\be \delta\left( W_{+2}^{(3)}[A] + W_{-2}^{(3)}[B] \right) = \int d^3 x \, J^{\mu\nu\lambda}
\p_{\nu}\omega_{\lambda\mu} \label{37} \ee

\no where now, compare with (\ref{14}), the Noether current
contains first and second derivatives terms, i.e.,

\be J^{\mu\nu\lambda} = - C^{\mu\nu\lambda}_{\s\rho\beta\gamma}
\p^{\beta} \tilde{g}^{\gamma\rho} - \frac 12
D^{\nu\lambda\mu\gamma\rho} \tilde{f}_{\gamma\rho} \label{38}, \ee

\no where $C^{\mu\nu\lambda}_{\s\rho\beta\gamma} $ is defined as
in (\ref{c}) while

\bea \tilde{g}_{\mu\nu} &=& A_{\mu\nu} + \tilde{\alpha} B_{\mu\nu}
\label{gtil} \\
\tilde{f}_{\mu\nu} &=& \frac{A_{\mu\nu}}{m_+} - \tilde{\alpha}
\frac{B_{\mu\nu}}{m_-} \label{ftil} \\
 D^{\nu\lambda\mu\gamma\rho}
&=& \epsilon^{\beta\lambda\nu}\left( 2 E_{\beta}^{\s\gamma}
E^{\rho\mu} - E_{\beta}^{\s\mu} E^{\rho\gamma} - \Box
\theta^{\gamma\rho} \eta^{\mu}_{\s\beta} \right) \label{39}. \eea

\no Introducing auxiliary fields which transform as $\delta
H_{\mu\nu\lambda} = - \p_{\mu} \omega_{\nu\lambda}$ we deduce

\be \delta\left( W_{+2}^{(3)}[A] + W_{-2}^{(3)}[B]  + \int d^3 x
\, J^{\mu\nu\lambda}H_{\mu\nu\lambda} \right) = \int d^3 x \,
\delta J^{\mu\nu\lambda}H_{\mu\nu\lambda}.\ee

\no However, we have now

\be  \delta J^{\mu\nu\lambda} = (1 +
\tilde{\alpha}^2)C^{\mu\nu\lambda}_{\rho\beta\gamma}\p^{\beta}\omega^{\gamma\rho}
- \frac 12 D^{\nu\lambda\mu\gamma\rho}\left( \frac 1{m_+} -
\frac{\tilde{\alpha}^2}{m_-}\right) \omega_{\gamma\rho} \label{40}.
\ee

\no As in the last section, we suppress the last term above by
fixing $\tilde{\alpha}$ up to a sign

\be \tilde{\alpha} = \pm \sqrt{\frac{m_-}{m_+}} \quad .
\label{alpha3} \ee

\no Consequently

\bea \delta J^{\mu\nu\lambda} &=& -(1 +
\tilde{\alpha}^2)C^{\mu\nu\lambda}_{\rho\beta\gamma}\p^{\beta}\omega^{\gamma\rho}
 \label{deltaj3}\\
&=& (1 + \tilde{\alpha}^2)C^{\mu\nu\lambda}_{\rho\beta\gamma}
\delta H^{\rho\beta\gamma} \label{deltaj3b}. \eea

\no Note the sign difference to  (\ref{20}) and (\ref{21}). This
is  due to the ``wrong'' sign of the Einstein-Hilbert term in
(\ref{wmais3}) and (\ref{wmenos3}). Thus, after elimination of the
auxiliary fields we have, compare with (\ref{23}),

\be W_S^{(4)} = W_{+2}^{(3)}[A] + W_{-2}^{(3)}[B]  + \int d^3 x \,
\frac{\left\lbrack J^{*}_{\mu\nu}J^{*\nu\mu} -
(J^*)^2\right\rbrack } {8(1+\tilde{\alpha}^2)} \quad , \label{44}
\ee

\no where now

\be J_{\mu\nu}^{*} =
\epsilon_{\mu}^{\s\gamma\lambda}J_{\nu\gamma\lambda} = 2\,
T_{\mu\nu}(\tilde{g}) - \eta_{\mu\nu} T (\tilde{g}) - V_{\mu\nu}
\label{j*3} \ee

\no with

\bea V^{\mu\nu} &=& - \frac 12 \epsilon^{\mu}_{\s\gamma\rho} \,
D^{\gamma\rho\nu\epsilon\delta} \tilde{f}_{\epsilon\delta}
\label{v} \\
&=& \left( 2 E^{\mu\epsilon} E^{\delta\nu} +
E^{\mu\nu}E^{\epsilon\delta} - \Box \eta^{\mu\nu}
\theta^{\epsilon\delta} \right) \tilde{f}_{\epsilon\delta}
\label{v2}. \eea

\no Rewriting the fields in term of the soldering invariant
combination $h_{\mu\nu}= \tilde{\alpha} A_{\mu\nu} - B_{\mu\nu}$:

\bea \tilde{f}_{\mu\nu} &=& \frac{A_{\mu\nu}}{m_+} -
\tilde{\alpha} \frac{B_{\mu\nu}}{m_-} = \frac{\tilde{\alpha}}{m_-}
h_{\mu\nu}
\label{49}\\
g_{\mu\nu} &=& A_{\mu\nu} + \tilde{\alpha}B_{\mu\nu} =
\tilde{\alpha} \, A_{\mu\nu} - h_{\mu\nu} \label{50}. \eea

\no It turns out that $W_S^{(4)}$ only depends on $h_{\mu\nu}$. In
particular, all the fourth-order terms in $W_S^{(4)}$ stem from
the combination:

\be \int \frac{d^3 x}{8(1+\tilde{\alpha}^2)}\left( V_{\mu\nu}
V^{\nu\mu} - V^2 \right) = \int \frac{d^3
x}{4(1+\tilde{\alpha}^2)} h_{(\mu\nu)}\left(\frac{2\,
\theta^{\mu\epsilon} \theta^{\nu\delta}-
\theta^{\mu\nu}\theta^{\epsilon\delta}}{m_+ m_-}\right) \Box^2
h_{(\epsilon\delta)} \nn \\
\label{52}. \ee

\no In deriving (\ref{52}) from (\ref{v2}) we have used
integration by parts, the identities
$E_{\mu\gamma}E^{\gamma}_{\s\nu} = \Box \theta_{\mu\nu} \,$ ; $\,
E_{\mu\nu}\theta^{\nu\gamma} = E_{\mu}^{\s\gamma} \, $ ; $ \,
\theta_{\mu\nu}\theta^{\nu\gamma} = \theta_{\mu}^{\s\gamma}$,
equation (\ref{49}) and $\left(\tilde{\alpha}/m_-\right)^2 =
1/m_+m_-$. After collecting all terms in (\ref{44}) we can write
the corresponding soldered Lagrangian density $\cL_{S}^{(4)}$ as:

\be \cL_{S}^{(4)} = \frac{1}{2(1+\tilde{\alpha}^2)}
h_{(\mu\nu)}\left\lbrack E^{\epsilon\mu}E^{\nu\delta} +
\frac{4(m_+-m_-)}{m_+ m_-} E^{\mu\delta}\Box \theta^{\nu\delta} +
\frac{\left( 2
\theta^{\mu\epsilon}\theta^{\nu\delta}-\theta^{\mu\nu}\theta^{\epsilon\delta}\right)\Box^2}{2
m_+ m_-}\right\rbrack h_{(\epsilon\delta)} \label{53}. \ee

\no Thus, as in the last section, the requirement of invariance
under local shifts proportional to each other effectively solders
the fields $A_{\mu\nu}$ and $B_{\mu\nu}$ into one combination
$h_{\mu\nu} = \tilde{\alpha} A_{\mu\nu} - B_{\mu\nu}$. By using
$\tilde{\alpha}=\pm\sqrt{m_-/m_+}$, it is easy to check that each
of the terms of (\ref{53}) is invariant under the discrete
symmetry $(m_+,m_-,\gamma_+,\gamma_-) \to
(-m_-,-m_+,\gamma_-,\gamma_+)$ which interchanges $W_{\pm 2}^{(3)}
\rightleftharpoons W_{\mp 2}^{(3)}$. Furthermore, the action
(\ref{53}) is invariant under the local symmetries $\delta
h_{\mu\nu} = \p_{\mu}\xi_{\nu} +
\epsilon_{\mu\nu\lambda}\Lambda^{\lambda}$ inherited from
$W_{\pm}^{(3)}$. The three terms in (\ref{53}) correspond exactly
to the quadratic truncation of the new massive gravity of
\cite{bht} up to an overall constant:

\be 2(1+\tilde{\alpha}^2)\cL_{S}^{(4)} = \left\lbrack - \sqrt{-g}R
 + \frac{m_+ - m_-}{m_+
m_-}\epsilon^{\mu\nu\rho}\Gamma_{\mu\gamma}^{\epsilon}\p_{\nu}\Gamma_{\epsilon\rho}^{\gamma}
+ \frac{\sqrt{-g}}{m_+ m_-}\left(R_{\mu\nu} R^{\nu\mu} - \frac 38
R^2 \right) \right\rbrack_{hh} \label{bht}. \ee

\no In identifying (\ref{53}) with (\ref{bht}) we have used
$g_{\mu\nu}= \eta_{\mu\nu} + 2 h_{(\mu\nu)}$ (or
$e_{\mu\nu}=\eta_{\mu\nu} + h_{\mu\nu}$). The second term in
(\ref{bht}) is the quadratic truncation of the topologically
massive gravity of \cite{djt}: $\cL_{TPM} =
\epsilon^{\mu\nu\rho}\Gamma_{\mu\gamma}^{\epsilon}\left\lbrack
\p_{\nu}\Gamma_{\epsilon\rho}^{\gamma} +
(2/3)\Gamma_{\nu\delta}^{\gamma}\Gamma_{\rho\epsilon}^{\delta}
\right\rbrack $.

\section{Conclusion}

In both cases of spin 1 and 2 theories in $D=2+1$ we have shown
that the simple addition of two first-order self-dual models
(parity singlets) of opposite helicities leads us to a parity
invariant theory (for equal masses) which describes a parity
doublet by means of a single field. Those are the well known Proca
and Fierz-Pauli theories respectively. On the other hand, the
addition of self-dual models with gauge symmetry demands extra
interference terms between the opposite helicity states in order
to produce the desired result. We have shown here that the
generalized soldering furnishes those required terms in a
systematic way also for spin 2 particles in a complete off-shell
procedure. In particular, the Fierz-Pauli theory with its
nontrivial mass term is automatically produced, section 3, out of
two second-order self-dual models of opposite helicities which are
the spin 2 analogues of the Maxwell-Chern-Simon theories.

In section 3 we have shown that if we start with two spin 2 self-dual models of third-order (quadratic
truncation of topologically massive gravity) we end up, after soldering, exactly with the new massive gravity
theory of \cite{bht}. Since in previous examples \cite{gsd}, the theories related via soldering turn out to be
equivalent (up to contact terms in the correlation functions), our results suggest that there might be a local
dual map between the gauge invariant sectors of $W_{+ 2}^{(3)} + W_{- 2}^{(3)}$ and the new massive gravity
theory \cite{bht} at linearized level. In particular, both theories have the same $m\to\infty$ limit (pure
Einstein-Hilbert) contrary to the Fierz-Pauli theory (see discussion in \cite{deser1}).

Extensions of the soldering formalism beyond the linear level in $D=2+1$ as well as the introduction of
interactions in the soldering of the third-order self-dual models are currently under investigation both in the
soldering and master action approaches. Moreover, it would be interesting to investigate, see also
\cite{deser3,khoudeir3}, higher spin ($s\ge 3$) generalizations of the soldering procedure in $D=2+1$ and their
possible relationships with massless higher spin theories in $D=3+1$.

\section{Acknowledgements}

D.D. is partially supported by \textbf{CNPq} while E.L.M. is
supported by \textbf{CAPES}. We thank Alvaro de Souza Dutra and
Marcelo Hott for discussions. We also thank A.Khoudeir for drawing
our attention to \cite{aak}.


\begin{thebibliography}{99}

\bibitem{stone}  M. Stone, Illinois preprint, ILL - (TH) - 89-23, 1989; Phys.
Rev. Lett. \textbf{63} (1989) 731; Nucl. Phys. B \textbf{327} (1989) 399

\bibitem{amorim}  R. Amorim, A. Das and C. Wotzasek, Phys. Rev. D \textbf{53}
(1996) 5810.

\bibitem{abw509}  E. M. C. Abreu, R. Banerjee and C. Wotzasek, Nucl. Phys.
B \textbf{509} (1998) 519.

\bibitem{djt} S. Deser, R. Jackiw and S. Templeton, Ann. of Phys. {\bf 140}(1982)
372.

\bibitem{bk}  R. Banerjee and S. Kumar, Phys. Rev. D \textbf{60}, (1999)
085005.

\bibitem{gsd} D. Dalmazi, A. de Souza Dutra and E. M. C. Abreu,  Phys.Rev.D {\bf 74}025015 (2006);
[Erratum-ibid. Phys. Rev. D\textbf{79} 109902 (2009)].

\bibitem{fierz} M. Fierz, Helv. Phys. Acta {\bf 12} (1939) 3; M.
Fierz, W. Pauli, Proc. Roy. Soc. {\bf 173} (1939) 211.

\bibitem{ilha} A. Ilha and C. Wotzasek, Phys. Rev. D {\bf 63},
(2001) 105013.

\bibitem{tpn} P.K. Townsend, K. Pilch and P. van Nieuwenhuizen, Phys. Lett B {\bf 136} (1984)38.

\bibitem{aragone} C. Aragone and A. Khoudeir, Phys. Lett.
B{\bf173} 141 (1986).

\bibitem{desermc} S. Deser and J. McCarthy, Phys. Lett. B{\bf 246}
441 (1990).

\bibitem{bht} E. Bergshoeff, O. Hohm and P.K. Townsend, ``Massive Gravity in Three
Dimensions'', arXiv:0905.1259 [hep-th].

\bibitem{oda} M. Nakazone and I. Oda, ``On Unitarity of Massive Gravity in Three Dimensions.
'', arXiv:0902.3531 [hep-th].

\bibitem{dj} S.Deser and R. Jackiw, Phys.Lett.B {\bf 139} (1984)
371.

\bibitem{gaitan}R. Gaitan, ``On the Coupling Problem of Higher Spin Fields in 2+1 Dimension '',
 PhD thesis, in spanish, arXiv:0711.2498.

\bibitem{prd2009} D.Dalmazi and E.L.Mendon\c ca, Phys.Rev.D\textbf{79} 045025 (2009).

\bibitem{aak}C. Aragone, P.J. Arias and A. Khoudeir, Nuovo Cim.B\textbf{109}, 303-310
(1994).

\bibitem{scaria} T. Scaria, ``Studies in certain planar field theories'', PhD Thesis, hep-th/0407022.

\bibitem{deser1} S. Deser, gr-qc/9211010.

\bibitem{deser3}T. Damour, S. Deser, Annales Poincare Phys.Theor. \textbf{47} 277 (1987).

\bibitem{khoudeir3} C. Aragone , A. Khoudeir,``Massive triadic Chern-Simons spin 3 theory'', Presented at 8th Latin
American Symposium on Relativity and Gravitation (SILARG), Sao Paulo, Brazil, 25-30 Jul 1993. Published in
SILARG 1993:0529-533, hep-th/9310048.


\end{thebibliography}
\end{document}